# Stability, Adsorption and Diffusion of $CH_4$, $CO_2$ and $H_2$ in Clathrate Hydrates

Guillermo Román-Pérez,* Mohammed Moaied, Jose M. Soler, and Felix Yndurain
*Departamento de Física de la Materia Condensada. Universidad Autónoma de Madrid. Cantoblanco. 28049 Madrid. Spain.*
(Dated: July 2, 2010)

We present a study of the adsorption and diffusion of $CH_4$, $CO_2$ and $H_2$ molecules in clathrate hydrates using *ab initio* van der Waals density functional formalism [Dion et al. Phys. Rev. Lett. **92**, 246401 (2004)]. We find that the adsorption energy is dominated by van der Waals interactions and that, without them, gas hydrates would not be stable. We calculate the maximum adsorption capacity as well as the maximum hydrocarbon size that can be adsorbed. The relaxation of the host lattice is essential for a good description of the diffusion activation energies, which are estimated to be of the order of 0.2, 0.4, and 1.0 eV for $H_2$, $CO_2$, and $CH_4$, respectively.

The existence of complex crystalline structures, made of water molecules hosting in their cavities hydrocarbons and other molecules, that stabilize the otherwise unstable network, has been known for many years [1–3]. These gas hydrates, or clathrates, are stable at high pressures and low temperatures. They are very abundant in the Earth's permafrost and marine sediments [4–6], and they have been detected in other planetary bodies like Mars and some moons of Saturn [7, 8]. They can be prepared in the laboratory under appropriate conditions, and different structural and spectroscopic measurements, like X-ray and neutron diffraction, nuclear magnetic resonance (NMR), Raman and infrared spectroscopy, and others, have been performed to characterize their composition and crystal structure (see [2] and references therein).

These compounds are important for many practical reasons. Historically, it was soon realized that their formation, in extraction and transportation pipes of hydrocarbons, had to be controlled to avoid their clogging. On a more global perspective, they could potentially be a huge source of hydrocarbons, with reserves much larger than those of oil and natural gas together [1]. However, their destabilization and release to the atmosphere, due to the temperature increase associated to global warming, constitutes a very serious environmental threat. Another source of interest is in their potential use for $H_2$ storage and also for $CO_2$ sequestration. On this respect, the extraction of $CH_4$ from natural hydrates, and its simultaneous substitution by $CO_2$, preserving their structure and stability, would be an ideal operation. The viability of such an operation requires, however, a precise knowledge of several magnitudes, like the relative stability of $CH_4$ and $CO_2$ hydrates, and the diffusion barriers of these molecules in the networks. We address these magnitudes in this work.

The crystalline structure of the clathrates is made up of H-bonded water molecules forming a network with cages of different shapes and sizes. Out of the various crystalline structures, we will address here the so called Structure I and Structure H (see Figure 1). For more details see for instance the Sloan's reviews [1–3] and references therein.

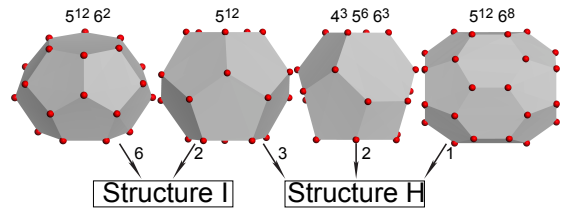

FIG. 1: (Color online) Basic units of the clathrates studied in this work. Structure I is formed by two $5^{12}$ cages and six $5^{12}6^2$ cages for a total of 46 $H_2O$ molecules per unit cell. Structure H is formed by three $5^{12}$ cages, two $4^35^66^3$ cages and one $5^{12}6^8$ cage, and 34 $H_2O$ molecules per unit cell. The red dots indicate the oxygen atom positions. Hydrogen atoms are not shown. The cages are not scaled with respect to each other

From a theoretical point of view, the study of these compounds represents a challenging task: it is necessary to deal, within the same methodology, with intramolecular covalent bonds, with the hydrogen bonds between $H_2O$ molecules, and with the van der Waals interaction between the host cages and the trapped molecules. Among the different questions we want to answer are: *i)* the adsorption energy and the relative stability of various molecules, like $CH_4$, $CO_2$, and $H_2$; *ii)* the maximum number of molecules that can be trapped inside each cavity, the interaction between molecules in different cavities and how these molecules stabilize the clathrates with respect to pure ice; *iii)* the maximum length of a hydrocarbon chain to be stably trapped; *iv)* the energy barrier for diffusion of different molecules.

There have been only a few *ab initio* calculations of gas hydrates. Petchkovskii and Tse [9], using different quantum chemistry methods that include dispersive forces, have studied the occupancy and thermodynamic stability of $H_2$ in isolated clathrate cages, with no relaxation. Alavi and Ripmeester [10] have calculated activation energy for $H_2$ diffusion, also using quantum chemistry methods in isolated cages. There have been several molecular dynamics (MD) simulations, using empirical potentials as well as *ab initio* density functional theory (DFT) [11, 12], to study the vibrational modes of the host molecules. MD simulations have been used also to study $H_2$ storage [13].

We have performed our DFT calculations with the SIESTA method [14, 15], which uses numerical basis sets for the valence electrons and norm-conserving pseudopotentials. For exchange and correlation we use the non local van der Waals density functional (vdW-DF) of Dion et al. [16] as implemented by Román-Pérez and Soler [17, 18]. We use a basis of optimized double-$\zeta$ orbitals including polarization orbitals (DZP) which have been successfully tested in water [20]. We have tested also the convergence of real-space and $k$-sampling integrations, and our residual forces were smaller than 5 meV/Å. All the calculations are for a single unit cell of the periodic solid, even though individual cages are shown for a better visualization.

First, using the vdW-DF, we have calculated the crystal parameters of structures I and H with no guest molecules inside. Although clathrates are only stable when occupied with molecules this is an appropriate starting point for our study (see below). For structure I we obtain a lattice parameter a of 12.09 Å to be compared with the experimental value [3] of 12.0 Å. For structure H the agreement with experiments [3] is also fair: $a = b = 12.37$ Å and $c = 10.24$ Å versus experimental values of $a = b = 12.2$ Å and $c = 10.1$ Å.

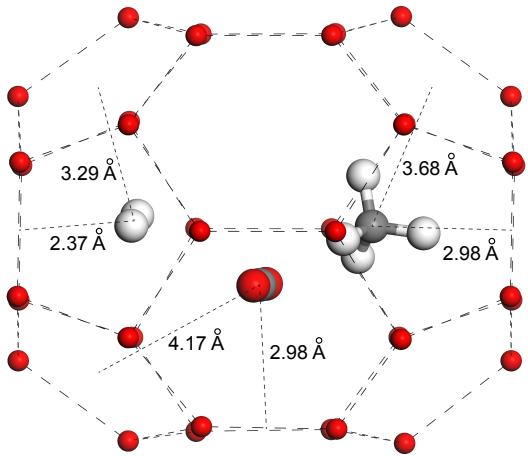

FIG. 2: (Color online) Equilibrium position of various molecules trapped in the $5^{12}6^8$ cavity of the clathrate structure H, obtained using the van der Waals density functional. The indicated positions are for individual molecules in the cavity. The relevant distances to the cage faces are indicated. The oxygen atoms of the cavity are indicated by small red circles. The broken lines serve as a guide to the eye.

We have calculated the adsorption energy of single $H_2$, $CO_2$ and $CH_4$ molecules in the different cavities of the host clathrates, including corrections for basis set superposition errors (BSSE) [21, 22]. We have used two different correlation functionals in order to assess the importance of the van der Waals contribution: the revPBE [19] generalized gradient approximation (GGA), that excludes van der Waals forces, and the vdW-DF. In Table I we show the results for the two different cavities of Structure I and the three different cavities of structure H. The results for $H_2$ can be compared with the value of 0.123 eV calculated by Patchkovskii and Tse [9] using quantum chemistry methods in isolated cavities.

TABLE I: Adsorption energies (in eV per molecule) for single $CH_4$, $CO_2$, and $H_2$ molecules in one of the different cavities of clathrate structures I and H. 'All' stands for one molecule inside each cavity. The adsorption energy is defined as the difference between the total energy of the clathrate, with the molecules inside, and that of the the empty clathrate plus the isolated molecules.

| Guest | Functional | Structure I | | | Structure H | | | |
|---|---|---|---|---|---|---|---|---|
| | | $5^{12}$ | $5^{12}6^2$ | All | $5^{12}$ | $4^35^66^3$ | $5^{12}6^8$ | All |
| $CH_4$ | GGA | 0.09 | -0.07 | -0.03 | 0.07 | 0.05 | -0.12 | 0.04 |
| $CH_4$ | vdW | -0.52 | -0.59 | -0.51 | -0.53 | -0.54 | -0.48 | -0.55 |
| $CO_2$ | GGA | 0.35 | 0.10 | 0.15 | 0.29 | 0.22 | 0.01 | 0.27 |
| $CO_2$ | vdW | -0.41 | -0.56 | -0.51 | -0.41 | -0.43 | -0.38 | -0.44 |
| $H_2$ | GGA | 0.01 | -0.01 | -0.02 | 0.03 | -0.01 | 0.00 | 0.03 |
| $H_2$ | vdW | -0.20 | -0.19 | -0.21 | -0.18 | -0.22 | -0.15 | -0.18 |

$CO_2$ and $CH_4$ have a similar adsorption energy of $\sim$0.5 eV, while that of $H_2$ is $\sim$0.2 eV. In all cases it is governed by the van der Waals interaction so that, if this is not included (i.e. in the GGA approximation), the adsorption is not energetically favorable. Therefore hereafter only the results with the vdW-DF will be considered. The adsorption energy of all the molecules is nearly independent of the structure and the cavity where they are adsorbed. Also, the last column of Table I shows that the interaction between different molecules adsorbed in different cavities is small. These results enable us to estimate the amount of guest molecules needed to stabilize the structure. We have calculated that ice, in the $Cmc2_1$ [23] structure, is energetically more stable than empty clathrates by 0.024 eV and 0.030 eV per $H_2O$ molecule for structures I and H respectively. Therefore, with the data of Table I, we can estimate that at least 3, 3 and 6 molecules of $CH_4$, $CO_2$ and $H_2$, respectively, are needed per unit cell, in order to energetically stabilize the clathrates with respect to empty ice and gas-phase molecules.

Figure 2 shows the equilibrium positions of the $H_2$, $CO_2$, and $CH_4$ molecules inside the $5^{12}6^8$ cavity of structure H (only one molecule at a time). In all cases, the water molecules are also relaxed. It is interesting to notice that none of the equilibrium positions is at the center of the cavity, indicating an attractive interaction with the water cage. This interaction has been shown to be responsible for the variation of the thermal conduction with temperature, from crystal-like to glass-like, in methane hydrates [24].

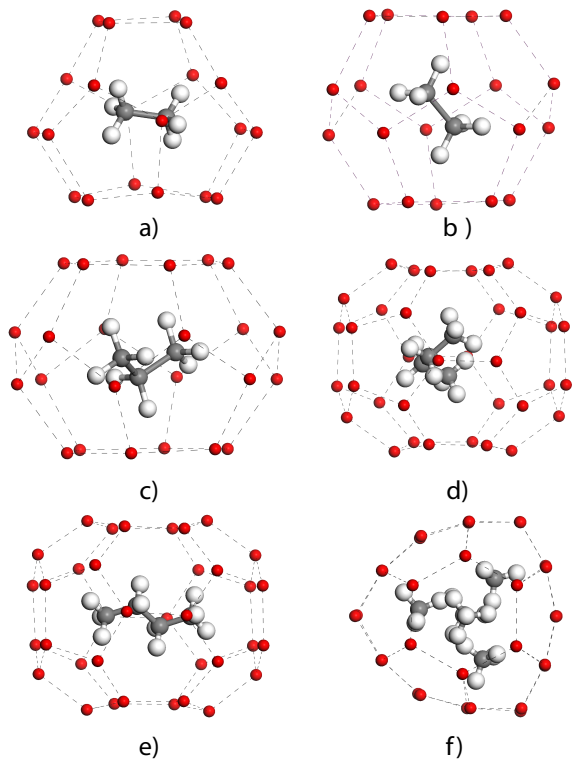

FIG. 3: (Color online) Hydrocarbon chains accommodated inside the different cavities of clathrates. a) Ethane ($C_2H_6$) at the $4^35^66^3$ cavity of structure H. b) Ethane at the $5^{12}$ cavity of structure H. c) Propane ($C_3H_8$) at the $5^{12}6^2$ cavity of the structure I. d) and e) Twisted ant elongated butane molecule ($C_4H_{10}$) at the $5^{12}6^8$ cavity of structure H. f) Five methane ($CH_4$) molecules inside the $5^{12}6^8$ cavity of structure H. The oxygen atoms of water molecules are indicated by small red circles. The hydrogen atoms are not shown. The broken lines are guides to the eye for a better visualization of the structure.

TABLE II: Incremental adsorption energies (work to adsorb each new molecule, in eV) of $CO_2$ and $CH_4$ molecules in the $5^{12}6^8$ cavity of structure H.

| Molecules | 1 | 2 | 3 | 4 | 5 | 6 | 7 |
|---|---|---|---|---|---|---|---|
| $CH_4$ | -0.48 | -0.42 | -0.48 | -0.30 | -0.17 | +0.35 | - |
| $CO_2$ | -0.38 | -0.37 | -0.47 | -0.31 | +0.05 | +0.10 | +1.27 |

TABLE III: Adsorption energies $\Delta E$ (in eV) of different hydrocarbons in various clathrate cavities.

| Molecule | Cavity | Host | $\Delta E$ (eV) |
|---|---|---|---|
| $C_2H_6$ | $4^35^66^3$ | H | -0.18 |
| $C_2H_6$ | $5^{12}$ | H | -0.24 |
| $C_3H_8$ | $5^{12}$ | H | +0.54 |
| $C_3H_8$ | $4^35^66^3$ | H | +0.24 |
| $C_3H_8$ | $5^{12}6^2$ | I | -0.37 |
| Twisted $C_4H_{10}$ | $5^{12}6^2$ | I | +0.06 |
| Twisted $C_4H_{10}$ | $5^{12}6^8$ | H | -0.76 |
| Elongated $C_4H_{10}$ | $5^{12}6^2$ | I | +0.33 |
| Elongated $C_4H_{10}$ | $5^{12}6^8$ | H | -0.86 |

Concerning the adsorption capacity, we find that, in all the cavities but the $5^{12}6^8$ cavity of structure H, only one $CH_4$ or $CO_2$ molecule can be accommodated. The results of filling up the $5^{12}6^8$ cavity of structure H, with the other cages empty, are given in Table II. We find a similar behavior for $CO_2$ and $CH_4$: the maximum energy gain per molecule corresponds to 3 molecules inside the cavity. The maximum capacity is five molecules for $CH_4$ and four for $CO_2$. These results are essentially independent of whether the adjacent cavities are occupied or not. Since the interaction between the molecules and their cages is expected to be large [24], all atoms are allowed to relax and we find that indeed the relaxation of the host lattice is essential to obtain reliable results. Thus, the ability of the relaxed $5^{12}6^8$ cage to accommodate molecules is much larger than simple geometrical arguments would indicate [1].

Another important information to assess the ability of clathrates to incorporate hydrocarbons is the maximum molecular size that they can accommodate. Table III reports the energy gain to incorporate hydrocarbon molecules of various sizes in different cavities, and Figures 3(a-e) show their relaxed geometries. Also, we show in Figure 3(f) the remarkably deformed geometry of the $5^{12}6^8$ cavity of structure H, saturated with 5 $CH_4$ molecules. This is similar to the capacity of porous metal-organic framework-5 (MOF-5) to incorporate hydrogen [25, 26]. From the data of Table III we conclude that $C_4H_{10}$ alone is not enough to stabilize energetically the clathrate with respect to ice.

The diffusion of guest molecules through the clathrate solid is an essential information concerning their possible storage and extraction. The calculated energy barriers required for the molecules to pass from one clathrate cavity to a neighbor one are shown in Figure 4. The relaxation of the host lattice is of paramount importance, and we observe that the energy barrier depends strongly on the molecule. For $H_2$, we obtain a barrier of 0.28 eV, close to the value of 0.250-0.283 eV calculated by Alavi and Ripmeester [10] using quantum chemistry methods in unrelaxed isolated cages. The diffusion of $H_2$ has been measured using NMR [27] obtaining a high diffusivity and an estimated activation energy of about 0.03 eV. Although our calculated value can be considered as an upper bound, since we assume that hydrogen does not share the cavity with other molecules, the discrepancy with the experimental data indicates the need of further study. The barrier for methane is much larger (1.17 eV) and it entails a substantial relaxation of the host structure. The unrelaxed H-bond length (1.82 Å) of the host hexagon becomes 1.89, 1.98 and 2.15 Å when the $H_2$, $CO_2$ and $CH_4$ molecules, respectively, pass through it. The energy barrier (0.42 eV) for $CO_2$ is smaller than that of $CH_4$, indicating a higher diffusivity. We have also studied the possible diffusion through the smaller pentagon face of the $5^{12}6^8$ cavity of structure H. We find, in addition to much larger barriers, that forcing the molecules to pass through the pentagon destroys the clathrate structure.

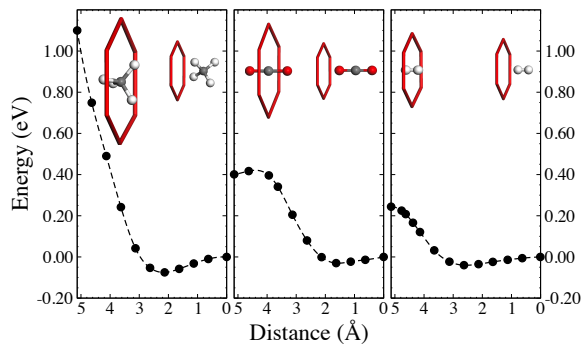

FIG. 4: (Color online) Total energy (with respect to its value at the cavity center) of a single molecule along the line connecting the center of an hexagonal face of the $5^{12}6^8$ cavity, in clathrate structure H. The relaxation (not scaled) of the hexagon due to the presence of the molecule is sketched to indicate a much larger relaxation in the $CH_4$ case (see text).

In summary, we have shown that first principles calculations, with the appropriate density functional to include van der Waals forces, can properly account for the stability of $CH_4$, $CO_2$ and $H_2$ water clathrates. We find that methane and carbon dioxide have a similar adsorption energy, higher than that of hydrogen. These results will enable us to study in the future the formation process of the clathrates [28, 29] as well as the phase diagram concerning their stability, as a function of pressure and temperature. Work in these directions is in progress and will be reported elsewhere.

We would like to thank Prof. E. Artacho for many lively and illuminating discussions. We also acknowledge grants FIS2009-12712 and CSD2007-00050 from the Spanish Ministry of Science and Innovation.

* Electronic address: guillermo.roman@uam.es.
[1] E. D. Sloan, Nature **426**, 353 (2003).
[2] A.K. Sum, C.A. Koh and E.D. Sloan, Industrial and Engineering Chemistry Research **48**, 7457 (2009).
[3] E. D. Sloan and C. A. Koh, "Clathrate Hydrates" CRC Press Taylor and Francis Goup (2008).
[4] T. S. Collett and M. W. Lee, Annals of the New York Academy of Sciences **912**, 51 (2000).
[5] A.G. Judd, M. Hovland, L.I. Dimitrov, S. Garcia and V. Jukes, Geofluids **2**, 109 (2002).
[6] M. D. Max, "Natural Gas Hydrate in Oceanic and Permafrost Environments" (Kluwer Academics Publishers, 2003).
[7] V.C. B. K. Chastain, Planetary and Space Science **55**, 1246 (2007).
[8] J. S. Loveday, R.J. Nelmes, M. Guthrie, S.A. Belmonte, D.R. Allan, D.D. Klug, J.S. Tse and Y. P. Handa, Nature **410** 661 (2001).
[9] S. Patchkovskii and J. S Tse, Proceedings of the National Academy of Sciences **100**, 14645 (2003).
[10] S. Alavi, J.A. Ripmeester, Angew. Chem. **2007**, 119, 6214 (2007).
[11] J. S. Tse, Journal of Supramolecular Chemistry **2**, 429 (2002).
[12] J. Wang, H. Lu and J. A. Ripmeester, J. Am. Chem. Soc. **131**, 40 (2009)
[13] H. Endou, K. Makino, H. Iwamoto, Y. Koba and M. Nakano, Molecular Dynamic Simulations of Hydrogen Storing in Clathrate Hydrates. In Proceedings of the 6th International Conference on Gas Hydrates (ICGH 2008), Vancouver, British Columbia, Canada, July 6-10, 2008.
[14] J. M. Soler, E. Artacho, J.D. Gale, A. Garcia, J. Junquera, P. Ordejón, and D. Sánchez-Portal, J. Phys.: Condens. Matter **14**, 2745 (2002).
[15] P. Ordejón, E. Artacho and J.M. Soler, Phys. Rev. B **53**, R10441 (1996).
[16] M. Dion, H. Rydberg, E. Schroder, D.C. Langreth and B.I. Lundqvist, Phys. Rev. Lett. **92**, 246401 (2004).
[17] G. Román-Pérez and J. M. Soler, Phys. Rev. Lett. **103**, 096102 (2009).
[18] L. Kong, G. Román-Pérez, J. M. Soler and D. C. Langreth Phys. Rev. Lett. **103**, 0961023 (2009).
[19] Y. Zhang and W. Yang, Phys. Rev. Lett. **80**, 890 (1998).
[20] M.V. Fernández-Serra and E. Artacho, Phys. Rev. Lett. **96**, 016404 (2006).
[21] S. F. Boys and F. Bernardi, Molecular Physics **19**, 553 (1970).
[22] S. Simon, M. Duran and J.J. Dannenberg, Journal of Chemical Physics **105**, 11024 (1996).
[23] S. W. Rick, Journal of Chemical Physics, **122**, 0945041 (2005).
[24] N. J. English and J. S. Tse, Phys. Rev. Lett. **103**, 015901 (2009).
[25] M. Mattesini, J.M. Soler and F. Yndurain, Phys. Rev. B **73**, 094111 (2006).
[26] L. Kong, V. R. Cooper, N. Nijem, K. Li, Y. J. Chabal and D. C. Langreth, Phys. Rev. B **79**, 081407(R) (2009).
[27] T. Okuchi, I. L. Moudrakovski and J.A. Ripmeester, Appl. Phys. Lett. **91**, 171903 (2007).
[28] F. Lehmkühler, M. Paulus, C. Sternemann, D. Lietz, F. Venturini, C. Gutt and M. Tolan, J. Am. Chem. Soc. **131**, 585 (2009).
[29] M. R. Walsh, C. A. Koh, E. D. Sloan, A. K. Sum and D. T. Wu, Science **326**, 1095 (2009).